\documentclass[aps,print,showpacs,a4paper,12pt,onecolumn]{revtex4}
\usepackage{amssymb}
\usepackage{amsmath}
\usepackage{graphicx}

\input{tcilatex}
\begin{document}

\title{Dark Soliton Interaction of Spinor Bose-Einstein Condensates in an
Optical Lattice}
\author{Zai-Dong Li, and Qiu-Yan Li}
\affiliation{Department of Applied Physics, Hebei University of Technology, Tianjin
300130, China}

\begin{abstract}
We study the magnetic soliton dynamics of spinor Bose-Einstein condensates
in an optical lattice which results in an effective Hamiltonian of
anisotropic pseudospin chain. An equation of nonlinear Schr\"{o}dinger type
is derived and exact magnetic soliton solutions are obtained analytically by
means of Hirota method. Our results show that the critical external field is
needed for creating the magnetic soliton in spinor Bose-Einstein
condensates. The soliton size, velocity and shape frequency can be
controlled in practical experiment by adjusting\textbf{\ }the magnetic
field. Moreover, the elastic collision of two solitons is investigated in
detail.
\end{abstract}

\maketitle

\section{Introduction}

The realization of spinor Bose-Einstein condensates (BECs) opens a useful
tool to understand and to confirm the dynamics of periodic structures in
solid state physics. Recently, the research of spinor BECs trapped in
optical potentials have received much attention both experimentally \cite%
{Stenger,Anderson} and theoretically \cite{Ho}. Due to the internal degrees
of freedom for the hyperfine spin of the atoms, spinor BECs bring forth a
rich variety of phenomena such as spin domains \cite{Miesner,Gemelke} and
textures \cite{Ohmi}. When the potential valley is so deep that the
individual sites are mutually independent, spinor BECs at each lattice site
behave like spin magnets and can interact with each other through both the
light-induced and the static, magnetic dipole-dipole interactions. These
site-to-site dipolar interactions can cause the ferromagnetic phase
transition \cite{Pu,Kevin} leading to a \textquotedblleft
macroscopic\textquotedblright\ magnetization of the condensate array, the
spin-wave like excitation \cite{Pu,Kevin,Zhang,Rigol} and magnetic soliton 
\cite{Xie,Lizd} analogous to the spin-wave and magnetic soliton in a
ferromagnetic spin chain. For a practical spin chain, the site-to-site
interaction is caused mainly by the exchange interaction, while the
dipole-dipole interaction is negligibly small. For the spinor BECs in the
optical lattice, the exchange interaction is absent. The individual spin
magnets are coupled by the magnetic and the light-induced dipole-dipole
interactions \cite{Zhang,Rigol} which are no longer negligible due to the
large number of atoms $N$ at each lattice site, typically of the order of
1000 or more. Therefore, the spinor BECs in an optical lattice offer a
totally new environment to study spin dynamics in periodic structures. The
magnetic soliton excited by the interaction between the spin waves is an
important and interesting phenomenon in spinor BECs. The Heisenberg model of
spin-spin interactions plays a significant role in understanding many
complex magnetic structures in solids. It explains the existence of
ferromagnetism and antiferromagnetism at temperatures below the Curie
temperature. The magnetic soliton \cite{Kosevich}, which describes localized
magnetization, is an important nonlinear excitation in the Heisenberg model 
\cite{Tjon,Li,Ablowitz,Huang}. However, the generation of controllable
solitons is an extremely difficult task due to the complexity of the
conventional magnetic materials. The spinor BECs seems an ideal system to
serve as a new test ground for studying the nonlinear excitations of spin
waves both theoretically and experimentally.

In this paper, we demonstrate that the magnetic soliton and elastic soliton
collision are admitted for spinor BECs in a one-dimensional optical lattice.
By means of Hirota method we obtain the analytical dark soliton solutions,
and also discuss the soliton interaction in detail. The outline of this
paper is organized as follows: In Sec. II the Solitons in a Spin Chain of
Atomic BEC's is investigated in detail. Next, we obtain the one-soliton
solution of spinor BECs in an optical lattice. In Sec. VI, the general
two-soliton solution is obtained. Analysis reveals that elastic soliton
collision occurs and there is a phase exchange during collision. Finally,
our concluding remarks are given in Sec. V.

\section{Solitons in a Spin Chain of Atomic BEC's}

The Hamiltonian describing an $F=1$ spinor condensate trapped in an optical
lattice, which is subject to the magnetic dipole-dipole interactions $H_{d}$
and is coupled to an external magnetic field via the magnetic dipole
Hamiltonian $H_{B}$ has the form as \cite{Ho,Miesner,Ohmi,Pu}%
\begin{equation}
H=H_{0}+H_{d}+H_{B},  \label{hamilton}
\end{equation}%
where the first term $H_{0}$ describes the interaction of the atoms with the
lattice potential $U_{L}\left( \mathbf{r}\right) $ and the ground-state
collisions which can be written as%
\begin{eqnarray*}
H_{0} &=&\sum_{\alpha }\int d\mathbf{r}\hat{\psi}_{\alpha }^{\dagger }(%
\mathbf{r})[-\frac{\hbar ^{2}\nabla ^{2}}{2m}+U_{L}(\mathbf{r})]\hat{\psi}%
_{\alpha }(\mathbf{r}) \\
&&+\sum_{\alpha ,\beta ,\mu ,\nu }\frac{\lambda _{s}}{2}\int d\mathbf{r}\hat{%
\psi}_{\alpha }^{\dagger }(\mathbf{r})\hat{\psi}_{\beta }^{\dagger }(\mathbf{%
r})\hat{\psi}_{\beta }(\mathbf{r}^{\prime })\hat{\psi}_{\alpha }(\mathbf{r})
\\
&&+\sum_{\alpha ,\beta ,\mu ,\nu }\frac{\lambda _{a}}{2}\int d\mathbf{r}\hat{%
\psi}_{\alpha }^{\dagger }(\mathbf{r})\hat{\psi}_{\mu }^{\dagger }(\mathbf{r}%
)\mathbf{F}_{\alpha \beta }\cdot \mathbf{F}_{\mu \nu }\hat{\psi}_{\nu }(%
\mathbf{r})\hat{\psi}_{\beta }(\mathbf{r}),
\end{eqnarray*}%
where $\hat{\psi}_{\alpha }\left( r\right) $ is the field annihilation
operator for an atom in the hyperfine state $\left\vert f=1,m_{f}=\alpha
\right\rangle $, the indices $\alpha ,\beta ,\mu ,\nu =$ $0,\pm 1$ denote
the Zeeman sublevels of the ground state, the parameters $\lambda _{s}$ and $%
\lambda _{a}$ characterize the short-range spin-independent and
spin-changing $s$-wave collisions, respectively. When the optical lattice
potential is deep enough there is no spatial overlap between the condensates
at different lattice sites. Under this condition the tight-binding
approximation is reasonable that the atomic field operator can be expanded
as $\hat{\psi}\left( \mathbf{r}\right) =\tsum\nolimits_{n}$ $%
\tsum\nolimits_{\alpha =0,\pm 1}\hat{a}_{\alpha }\left( n\right) \phi
_{n}\left( \mathbf{r}\right) $, where $n$ labels the lattice sites, $\phi
_{n}(\mathbf{r})$ is the Hartree wave function of the condensate for the $n$%
th microtrap and the operators $\hat{a}_{\alpha }(n)$ satisfy the bosonic
commutation relations $[\hat{a}_{\alpha }(n),\hat{a}_{\beta }^{\dag
}(l)]=\delta _{\alpha \beta }\delta _{nl}$. It is assumed that all Zeeman
components share the same spatial wave function. If the condensates at each
lattice site contain the same\textbf{\ }number of atoms\textbf{\ }$N$, the
ground-state wave functions for different sites have the same form $\phi
_{n}\left( \mathbf{r}\right) =f\left( \mathbf{r}-\mathbf{r}_{n}\right) $. In
this case the dipole-dipole interaction potential $H_{d}$ in Eq. (\ref%
{hamilton}) can be expressed by%
\begin{equation*}
H_{d}=\sum_{i}\sum_{j\left( \neq i\right) }\frac{\mu _{0}}{4\pi \left\vert 
\mathbf{r}_{ij}\right\vert ^{3}}\left[ \overrightarrow{\mu }_{i}\cdot 
\overrightarrow{\mu }_{j}-3\left( \overrightarrow{\mu }_{i}\cdot \mathbf{%
\hat{r}}_{ij}\right) \left( \overrightarrow{\mu }_{j}\cdot \mathbf{\hat{r}}%
_{ij}\right) \right] ,
\end{equation*}%
where the parameter $\overrightarrow{\mu }_{i}=\gamma \mathbf{S}_{i}$ is the
magnetic dipole moment at site $i$, with $\mathbf{S}_{i}=\hat{a}_{\alpha
}^{\dag }(i)\mathbf{F}_{\alpha \beta }\hat{a}_{\beta }(i)$ being the total
angular momentum operator and $\gamma $ the gyromagnetic ratio, $\mathbf{r}%
_{i}$ denotes the coordinate of the $i$th site, $\mathbf{r}_{ij}=\mathbf{r}%
_{i}-\mathbf{r}_{j}$, $\mathbf{\hat{r}}_{ij}=\mathbf{r}_{ij}/\left\vert 
\mathbf{r}_{ij}\right\vert $, $\mu _{0}$ is the vacuum permeability. Also
the Zeeman energy $H_{B}$ can be described by%
\begin{equation*}
H_{B}=-\gamma \sum_{i}\mathbf{S}_{i}\cdot \mathbf{B},
\end{equation*}%
where $\mathbf{B}$ is the external magnetic field.

In this paper we consider a one-dimensional optical lattice along the $z$%
-direction, which is also chosen as the quantization axis. In the absence of
spatial overlap between individual condensates, and neglecting unimportant
constants, the effective spin Hamiltonian can be constructed \cite{Zhang,Pu}
as%
\begin{equation}
H=\sum_{i}\left[ \lambda _{a}^{\prime }\mathbf{S}_{i}^{2}+\sum_{j\left( \neq
i\right) }J_{ij}\mathbf{S}_{i}\cdot \mathbf{S}_{j}-3\sum_{j\left( \neq
i\right) }J_{ij}S_{i}^{z}S_{j}^{z}-\gamma \mathbf{S}_{i}\cdot \mathbf{B}%
\right]  \label{hamilton5}
\end{equation}%
which determine the dynamics of the magnetic dipole moment of spinor BECs in
an optical lattice, where the parameter $\lambda _{a}^{\prime }=(1/2)\lambda
_{a}\int d^{3}r\left\vert \phi _{n}(\mathbf{r})\right\vert ^{4}$, here $%
\lambda _{a}$ characterizes the spin-dependent $s$-wave collisions, and $%
J_{ij}=\gamma ^{2}\mu _{0}/(4\pi \left\vert \mathbf{r}_{ij}\right\vert ^{3})$%
. The first term in Eq. (\ref{hamilton5})\textbf{\ }represent the
spin-dependent interatomic collisions at a given site, the second and the
third terms describe the site-to-site spin coupling induced by the static
magnetic field dipole-dipole interaction. Without this interaction terms,
the ground state of the Hamiltonian (\ref{hamilton5}) is $\left\vert
GS\right\rangle =\left\vert N,-N\right\rangle $ where $N=\sum_{i}N_{i}$ is
the total atomic number in the lattice. The total spin at site $i$ has the
expectation value $\left\langle \hat{S}_{i}^{z}\right\rangle =-N_{i}\hbar .$
Due to the large factor $N_{i}$, the magnetic dipole-dipole interaction in
the optical lattice can not be neglected. After the site-to-site coupling is
considered, the transfer of the transverse spin excitation from site to site
is allowed, resulting in the distortion of the ground state spin structure.
This distortion can propagate and hence generate magnetic soliton or spin
wave along the atomic spin chain.

It should be noted that the Holstein-Primakoff transformations \cite%
{Holstein} reported earlier is an useful tool to clear the physical meaning
of spin waves and magnetic soliton. For this transformation a local
spin-deviation operator is introduced firstly, i.e., $\hat{n}=\hat{S}-\hat{S}%
^{z}$ with the eigenvalues $n=S-m$ in which it can be seen that the
increasing $m$ decreases $n$ and vice versa. For the state $\left\vert
n\right\rangle $, the $\hat{S}^{-}\left\vert n\right\rangle =\sqrt{%
(2S-n)(1+n)}\left\vert n+1\right\rangle $ and $\hat{S}^{+}\left\vert
n\right\rangle =\sqrt{\left( 2S-n\right) \left( n-1\right) }\left\vert
n-1\right\rangle $, it can define the creation and annihilation operators $a$
and $a^{+}$ which satisfy the following boson commutator relation $[\hat{a},%
\hat{a}^{+}]=1,[\hat{a},\hat{a}]=[\hat{a}^{+},\hat{a}^{+}]=0.$ The result of 
$a$ and $a^{+}$ operate on the state $\left\vert n\right\rangle $ is $\hat{a}%
\left\vert n\right\rangle =\sqrt{n}\left\vert n-1\right\rangle $, $\hat{a}%
^{+}\left\vert n\right\rangle =\sqrt{n+1}\left\vert n+1\right\rangle $, and $%
\hat{a}^{+}\hat{a}\left\vert n\right\rangle =n\left\vert n\right\rangle $.
With the above relations, the spin operators $\hat{S}$ can be expressed by
the Bose operators $\hat{a}^{+}$ and $\hat{a}$, i.e., Holstein-Primakoff
transformations \cite{Holstein} as follows 
\begin{eqnarray}
\hat{S}^{+} &=&(\sqrt{2S-\hat{a}^{+}\hat{a}})\hat{a},  \notag \\
\hat{S}^{-} &=&\hat{a}^{+}(\sqrt{2S-\hat{a}^{+}\hat{a}}),  \notag \\
\hat{S}_{z} &=&(S-\hat{a}^{+}\hat{a}).  \label{Hp1}
\end{eqnarray}%
In fact the spin waves and magnetic soliton show the small distortion of the
ground-state spin structure, i.e., $n\ll S$. In this case we can expand the
Eq. (\ref{Hp1}) as 
\begin{eqnarray}
\hat{S}^{+} &=&\sqrt{2S}(1-\frac{\hat{a}^{+}\hat{a}}{4S}+\cdot \cdot \cdot )%
\hat{a},  \notag \\
\hat{S}^{-} &=&\hat{a}^{+}\sqrt{2S}(1-\frac{\hat{a}^{+}\hat{a}}{4S}+\cdot
\cdot \cdot ),  \notag \\
\hat{S}_{z} &=&(S-\hat{a}^{+}\hat{a}).  \label{Hp2}
\end{eqnarray}%
Substituting the Eq. (\ref{Hp2}) into the Hamiltonian (\ref{hamilton5}) and
keeping terms through fourth order in $\hat{a}$ and $\hat{a}^{\dagger }$, we
can get 
\begin{eqnarray}
H &=&N\lambda _{a}^{\prime }S\left( S+1\right) -\gamma NSB_{z}+\gamma
B_{z}\sum_{i}\hat{a}_{i}^{\dagger }\hat{a}_{i}+S\sum_{i}\sum_{j\left( \neq
i\right) }J_{ij}\left( \hat{a}_{i}\hat{a}_{j}^{\dagger }+\hat{a}%
_{i}^{\dagger }\hat{a}_{j}\right)  \notag \\
&&-\frac{1}{4}\sum_{i}\sum_{j\left( \neq i\right) }J_{ij}\left( \hat{a}%
_{j}^{\dagger }\hat{a}_{j}^{\dagger }\hat{a}_{j}\hat{a}_{i}+\hat{a}%
_{i}^{\dagger }\hat{a}_{i}\hat{a}_{i}\hat{a}_{j}^{\dagger }+\hat{a}%
_{i}^{\dagger }\hat{a}_{j}^{\dagger }\hat{a}_{j}\hat{a}_{j}+\hat{a}%
_{i}^{\dagger }\hat{a}_{i}^{\dagger }\hat{a}_{i}\hat{a}_{j}\right)  \notag \\
&&-2\sum_{i}\sum_{j\left( \neq i\right) }J_{ij}\left( S^{2}-S\left( \hat{a}%
_{j}^{\dagger }\hat{a}_{j}+\hat{a}_{i}^{\dagger }\hat{a}_{i}\right) +\hat{a}%
_{i}^{\dagger }\hat{a}_{i}\hat{a}_{j}^{\dagger }\hat{a}_{j}\right) .
\label{hamilton6}
\end{eqnarray}%
Under the spin coherent state and using the time-dependent variation
principle, the nonlinear operator motion equation of Hamiltonian (\ref%
{hamilton6}) can be transformed into an equation for the probability
amplitude $\psi _{k}=\langle \psi |a_{k}|\psi \rangle $ which describes the
nonlinear dynamics of coherent spin excitations on the lattice $k$.%
\begin{eqnarray}
i\hbar \frac{\partial }{\partial t}\psi _{k} &=&\left( \gamma
B_{z}+4JS\right) \psi _{k}+2JS\sum_{\delta }\psi _{k+\delta }-4J\sum_{\delta
}\psi _{k}\psi _{k+\delta }^{\ast }\psi _{k+\delta }  \notag \\
&&-\frac{1}{2}J\sum_{\delta }\left( 2\psi _{k}^{\ast }\psi _{k}\psi
_{k+\delta }+\psi _{k+\delta }^{\ast }\psi _{k+\delta }\psi _{k+\delta
}+\psi _{k}\psi _{k}\psi _{k+\delta }^{\ast }\right)  \label{NLS}
\end{eqnarray}%
where $\delta =\pm 1$, i.e., we consider only the nearest-neighbor
interactions which is a good approximation \cite{Konotop} for the BECs in a
one dimensional optical lattice as the large lattice constant.

\section{One soliton solution}

When the optical lattice is infinitely long and the spin excitations are in
the long-wavelength limit, $\psi _{k}$, $\psi _{k+\delta }\rightarrow \psi
\left( z,t\right) $ in the continuum limit approximation, we have 
\begin{equation}
-i\frac{\hbar }{J}\psi _{t}+\left( \frac{\gamma B_{z}}{J}+8S\right) \psi
+2S\psi _{zz}-12\psi \left\vert \psi \right\vert ^{2}=0.  \label{NLS1}
\end{equation}%
In fact, Eq. (\ref{NLS1}) has dark soliton solutions which can be obtained
by many methods. Here we take the Hirota bilinear transformation to get the
exact one- and two-soliton solutions of Eq. (\ref{NLS1}). For this method it
apply the direct transformation to the nonlinear equation which is the form 
\begin{equation}
\psi =\frac{g}{f},  \label{Hirota1}
\end{equation}%
where $g\left( z,t\right) $ are complex functions and $f\left( z,t\right) $
is a real function. Substituting (\ref{Hirota1}) into (\ref{NLS1}) we obtain 
\begin{equation}
-f\frac{\hbar }{J}\left( iD_{t}-\frac{2JS}{\hbar }D_{z}^{2}\right) g\cdot
f-4Sg\left[ D_{z}^{2}f\cdot f+\frac{3}{S}g\overline{g}-\left( \frac{\gamma
B_{z}}{4SJ}+2\right) f^{2}\right] =0,  \label{Hirota2}
\end{equation}%
where $D_{t}$ and $D_{z}^{2}$ are the Hirota bilinear operators which are
defined as 
\begin{equation}
D_{z}^{m}D_{t}^{n}g\left( z,t\right) \cdot f\left( z,t\right) =\left. \left( 
\frac{\partial }{\partial z}-\frac{\partial }{\partial z^{\prime }}\right)
^{m}\left( \frac{\partial }{\partial t}-\frac{\partial }{\partial t^{\prime }%
}\right) ^{n}g\left( z,t\right) f\left( z^{\prime },t^{\prime }\right)
\right\vert _{z=z^{\prime },t=t^{\prime }}.  \label{oper1}
\end{equation}%
Then Eq. (\ref{Hirota2}) can be decoupled as two equations 
\begin{eqnarray}
\hat{G}_{1}g\cdot f &=&0,\text{ }  \notag \\
\hat{G}_{2}f\cdot f &=&-\frac{3}{S}g\overline{g},  \label{Hirota3}
\end{eqnarray}%
where the Hirota bilinear operators $\hat{G}_{1}$ and $\hat{G}_{2}$ are
given by 
\begin{eqnarray}
\hat{G}_{1} &=&iD_{t}-\frac{2JS}{\hbar }D_{z}^{2}-\lambda ,\text{ }  \notag
\\
\hat{G}_{2} &=&D_{z}^{2}+\frac{\hbar \lambda }{4SJ}-\frac{\gamma B_{z}}{4SJ}%
-2,  \label{oper2}
\end{eqnarray}%
where $\lambda $ is constant to be determined. Now the Eq. (\ref{Hirota3})
has made the Eq. (\ref{NLS1}) to the normal procedure of Hirota method for
getting the exact soliton solutions. In the following by making a series of
suitable assumption for the expression of $g$ and $f$, the exact one- and
two-soliton solution can be obtained analytically. To this purpose we
suppose firstly that 
\begin{equation}
g=g_{0}\left( 1+\chi g_{1}\right) ,\text{ }f=1+\chi f_{1},  \label{ansaz1}
\end{equation}%
in order to construct a dark one-soliton solution for the system (\ref{NLS1}%
), where $\chi $ is an arbitrary parameter which is absorbed in expressing
the soliton solutions in the following sections. Substituting (\ref{ansaz1})
into (\ref{Hirota3}), and collecting the coefficients of $\chi ^{0}$, we
obtain 
\begin{eqnarray}
\hat{G}_{1}g_{0}\cdot 1 &=&0,\text{ }  \notag \\
\frac{\hbar \lambda }{4SJ}-\frac{\gamma B_{z}}{4SJ}-2 &=&-\frac{3}{S}g_{0}%
\overline{g}_{0},  \label{Hirota4}
\end{eqnarray}%
where the first equation in Eq. (\ref{Hirota4}) can be expressed by the
definition (\ref{oper1}) from which in is easy to find the solution
satisfying (\ref{Hirota4}) as 
\begin{equation}
g_{0}=\tau _{1}\exp \left( i\Phi _{1}\right) ,  \label{solution0}
\end{equation}%
where $\Phi _{1}$ has of form 
\begin{equation}
\Phi _{1}=\sqrt{\frac{\hbar }{2JS}}l_{1}z-\left( \lambda -l_{1}^{2}\right)
t-\Phi _{1}^{0},\text{ }  \label{solutionb2}
\end{equation}%
in which $l_{1}$, $\Phi _{1}^{0}$ are real constants and $\tau _{1}$ are
complex constant. With the restriction, i.e., the second equation in Eq. (%
\ref{Hirota4}) the expression of $\tau _{1}$ is 
\begin{equation}
\left\vert \tau _{1}\right\vert ^{2}=\frac{1}{12J}\left( \gamma
B_{z}+8SJ-\hbar \lambda \right)  \label{para1}
\end{equation}%
which imply the condition $\gamma B_{z}+8SJ-\hbar \lambda >0$, i.e., the
minimum external field value for the existence of soliton solution in Eq. (%
\ref{NLS1}), $B_{z}=\left( \hbar \lambda -8SJ\right) /\gamma $. The
coefficient of $\chi $ in Eq. (\ref{Hirota3}) leads to 
\begin{align}
\pounds _{1}\left( 1\cdot f_{1}+g_{1}\cdot 1\right) & =0,  \notag \\
\hat{G}_{2}\left( f_{1}\cdot 1+1\cdot f_{1}\right) & =-\frac{3}{S}\left\vert
\tau _{1}\right\vert ^{2}\left( g_{1}+\overline{g}_{1}\right) ,
\label{oper3}
\end{align}%
where $\pounds _{1}$ is defined as 
\begin{equation}
\pounds _{1}=iD_{t}-\frac{2JS}{\hbar }D_{z}^{2}-i2l_{1}\sqrt{\frac{2JS}{%
\hbar }}D_{z}.  \label{operatorb3}
\end{equation}%
By expanding Eq. (\ref{oper3}) with the definition (\ref{oper1}) in detail,
one can easily check that equations (\ref{oper3}) admit the following
solutions 
\begin{eqnarray}
g_{1} &=&Z_{g}\exp \xi _{1},\text{ }  \notag \\
f_{1} &=&\exp \xi _{1},  \label{solution1}
\end{eqnarray}%
where 
\begin{equation*}
\xi _{1}=P_{1}z-\Omega _{1}t-\xi _{1}^{0},
\end{equation*}%
in which $P_{1}$, $\Omega _{1}$and $\xi _{1}^{0}$ are real constants and $%
Z_{g}$ is a complex constant which is given by 
\begin{equation}
Z_{g}=-\frac{\left[ \frac{2JS}{\hbar }P_{1}^{2}-i\left( 2P_{1}l_{1}\sqrt{%
\frac{2JS}{\hbar }}+\Omega _{1}\right) \right] ^{2}}{\left( \frac{2JS}{\hbar 
}P_{1}^{2}\right) ^{2}+\left( 2P_{1}l_{1}\sqrt{\frac{2JS}{\hbar }}+\Omega
_{1}\right) ^{2}}.  \label{parab1}
\end{equation}%
From this equation it can easily be seen that $\left\vert Z_{g}\right\vert
^{2}=1$. Then using Eq. (\ref{ansaz1}), after absorbing $\chi $, the dark
one-soliton solution of spinor BECs in an optical lattice can be derived as 
\begin{equation}
\psi _{1}=\frac{g}{f}=\frac{\tau _{1}}{2}\left[ \left( 1+Z_{g}\right)
-\left( 1-Z_{g}\right) \tanh \frac{\xi _{1}}{2}\right] \exp \left( i\Phi
_{1}\right) .  \label{onesoliton1}
\end{equation}%
The solution in Eq. (\ref{onesoliton1}) describes a magnetic dark soliton
solution characterized by four real parameters: velocity of envelope motion $%
\Omega _{1}/P_{1}$, phase $\Phi _{1}$, coordinate of the center of the
solitary wave $\xi _{1}^{0}/P_{1}$ and initial phase $\sqrt{\frac{2JS}{\hbar 
}}\Phi _{1}^{0}/l_{1}$. This solution is similar to that of the Heisenberg
spin chain with an external field where the dipolar coupling is typically
several orders of magnitude weaker than the exchange coupling and would
correspond to Curie temperatures much below the observed ones. Hence its
contribution to the spin wave and magnetic soliton can be neglected in
practice. However, for the spinor BECs in the optical lattice the exchange
interaction is absent and the individual spin magnets are coupled by the
magnetic dipole-dipole interactions. Due to the large number of atoms $N$ at
each lattice site, these site to site interactions, despite the large
distance between sites, explain the natural existence of magnetic soliton
which agrees with the results in Refs. \cite{Pu,Zhang}. With the help of Eq.
(\ref{oper3}) we have 
\begin{equation}
\frac{S}{6P_{1}^{2}}=\frac{\left\vert \tau _{1}\right\vert ^{2}}{%
P_{1}^{4}+\left( 2P_{1}l_{1}\sqrt{\frac{\hbar }{2JS}}+\frac{\hbar }{2JS}%
\Omega _{1}\right) ^{2}}.  \label{parab2}
\end{equation}%
By combine the above presentation with the Eq. (\ref{para1}) we obtained the
following relation for $P_{1}\neq 0$%
\begin{equation*}
B_{z}=\frac{1}{\gamma }\left[ 2JSP_{1}^{2}+\left( 2l_{1}\sqrt{\hbar }+\frac{%
\hbar }{\sqrt{2JS}}\frac{\Omega _{1}}{P_{1}}\right) ^{2}-8SJ+\hbar \lambda %
\right] ,
\end{equation*}%
which shows that the external field $B_{z}$ and the dipole-dipole
interaction can affect the velocities, size and shape frequency of the
magnetic soliton. It offer an useful scheme to control soliton in practical
experiment by adjusting of\textbf{\ }the magnetic field.

\section{Dark soliton interaction}

In this section we will give the analytical expression of two soliton
solutions of Eq. (\ref{NLS1}). The properties is discussed in detail as
well. To this purpose for constructing the dark two-soliton solutions of Eq.
(\ref{NLS1}) we now assume that 
\begin{equation}
g=g_{0}\left( 1+\chi g_{1}+\chi ^{2}g_{2}\right) ,\text{ }f=1+\chi
f_{1}+\chi ^{2}f_{2},  \label{ansazb2}
\end{equation}%
where $g_{0}$ and $h_{0}$ are obtained here as in Eq. (\ref{solution0}). By
employing the same procedure before we obtain the following set of equations
from Eq. (\ref{Hirota3}), corresponding to the different powers of $\chi $
as follows

(i) for the coefficient of $\chi $%
\begin{align}
\pounds _{1}\left( 1\cdot f_{1}+g_{1}\cdot 1\right) & =0,  \notag \\
\hat{G}_{2}\left( f_{1}\cdot 1+1\cdot f_{1}\right) & =-\frac{3}{S}\left\vert
\tau _{1}\right\vert ^{2}\left( g_{1}+\overline{g}_{1}\right) ,
\label{Hirota5}
\end{align}%
where $\pounds _{1}$ is same with Eq. (\ref{operatorb3}).

(ii) for the coefficient of $\chi ^{2}$%
\begin{eqnarray*}
\hat{G}_{1}\left( g_{0}\cdot f_{2}+g_{0}g_{1}\cdot f_{1}+g_{0}g_{2}\cdot
1\right) &=&0, \\
\pounds _{1}\left( 1\cdot f_{2}+g_{1}\cdot f_{1}+g_{2}\cdot 1\right) &=&0,%
\text{ }
\end{eqnarray*}

\begin{equation}
\hat{G}_{2}\left( 1\cdot f_{2}+f_{2}\cdot 1+f_{1}\cdot f_{1}\right) =-\frac{3%
}{S}\left[ \left\vert \tau _{1}\right\vert ^{2}\left( g_{1}\overline{g}%
_{1}+g_{2}+\overline{g}_{2}\right) \right] .  \label{Hirota6}
\end{equation}

(iii) for the coefficient of $\chi ^{3}$%
\begin{align*}
\hat{G}_{1}\left( g_{0}g_{1}\cdot f_{2}+g_{0}g_{2}\cdot f_{1}\right) & =0, \\
\pounds _{1}\left( g_{1}\cdot f_{2}+g_{2}\cdot f_{1}\right) & =0,
\end{align*}%
\begin{equation}
\hat{G}_{2}\left( f_{1}\cdot f_{2}+f_{2}\cdot f_{1}\right) =-\frac{3}{S}%
\left\vert \tau _{1}\right\vert ^{2}\left( g_{1}\overline{g}_{2}+g_{2}%
\overline{g}_{1}\right) .  \label{Hirota7}
\end{equation}

(iii) for the coefficient of $\chi ^{4}$%
\begin{align*}
\hat{G}_{1}\left( g_{0}g_{2}\cdot f_{2}\right) & =0, \\
\pounds _{1}\left( g_{2}\cdot f_{2}\right) & =0,
\end{align*}%
\begin{equation}
\hat{G}_{2}\left( f_{2}\cdot f_{2}\right) =-\frac{3}{S}\left\vert \tau
_{1}\right\vert ^{2}g_{2}\overline{g}_{2}.  \label{Hirota8}
\end{equation}%
Repeating the same normal procedure in obtaining the one-soliton solution
one can easily get the solutions of the above set of equations as 
\begin{align}
g_{1}& =Z_{1}\exp \xi _{1}+Z_{2}\exp \xi _{2},\text{ }  \notag \\
\text{ }g_{2}& =A_{12}Z_{1}Z_{2}\exp \left( \xi _{1}+\xi _{2}\right) , 
\notag \\
f_{1}& =\exp \xi _{1}+\exp \xi _{2},  \notag \\
f_{2}& =A_{12}\exp \left( \xi _{1}+\xi _{2}\right) ,  \label{solution2}
\end{align}%
where%
\begin{equation*}
\xi _{j}=P_{j}z-\Omega _{j}t+\xi _{j}^{0},
\end{equation*}%
\begin{equation*}
Z_{j}=-\frac{P_{j}-i\sqrt{6\left\vert \tau _{1}\right\vert ^{2}/S-P_{j}^{2}}%
}{P_{j}+i\sqrt{6\left\vert \tau _{1}\right\vert ^{2}/S-P_{j}^{2}}},
\end{equation*}%
in which $j=1,2$. Noting that%
\begin{equation*}
\kappa _{j}=\left( \frac{6\left\vert \tau _{1}\right\vert ^{2}}{S}%
-P_{1}^{2}\right) ^{1/2}\frac{2JS}{\hbar },
\end{equation*}%
by which we can rewrite $Z_{j}$ and $\Omega _{j}$, $j=1,2$, as%
\begin{eqnarray*}
Z_{j} &=&-\frac{\frac{2JS}{\hbar }P_{j}-i\kappa _{j}}{\frac{2JS}{\hbar }%
P_{j}+i\kappa _{j}},\text{ } \\
\Omega _{j} &=&\left( \kappa _{j}-2l_{1}\sqrt{\frac{2JS}{\hbar }}\right)
P_{j},\text{ }
\end{eqnarray*}%
$\allowbreak $ $\allowbreak $So from Eq. (\ref{Hirota5}) to Eq. (\ref%
{Hirota8}) the expression of $A_{12}$ is obtained in the form%
\begin{equation*}
A_{12}=\frac{\left( P_{1}-P_{2}\right) ^{2}\left( \frac{2JS}{\hbar }\right)
^{2}+\left( \kappa _{1}-\kappa _{2}\right) ^{2}}{\left( P_{1}+P_{2}\right)
^{2}\left( \frac{2JS}{\hbar }\right) ^{2}+\left( \kappa _{1}-\kappa
_{2}\right) ^{2}}.
\end{equation*}%
Now using Eq. (\ref{Hirota1}), (\ref{ansazb2}) and (\ref{solution2}), the
dark two-solitons can be found explicitly as follows%
\begin{equation}
\psi _{two}=\frac{g}{f}=\tau _{1}\exp \left( i\Phi _{1}\right) \frac{%
1+Z_{1}\exp \xi _{1}+Z_{2}\exp \xi _{2}+A_{12}Z_{1}Z_{2}\exp \left( \xi
_{1}+\xi _{2}\right) }{1+\exp \xi _{1}+\exp \xi _{2}+A_{12}\exp \left( \xi
_{1}+\xi _{2}\right) },  \label{two}
\end{equation}%
where all the parameters have been obtained in the previous section. The
solution (\ref{two}) describes a general elastic scattering process of two
solitary waves with different center velocities $\Omega _{1}/P_{1}$ and $%
\Omega _{2}/P_{2}$ and the same phase $\Phi _{1}$ . Before collision, they
move towards each other, one with velocity $\Omega _{1}/P_{1}$ and the other
with $\Omega _{2}/P_{2}$. In order to understand the nature of two-soliton
interaction, we analyze the asymptotic behavior of two-soliton solution (\ref%
{two}). Asymptotically, the two-soliton waves (\ref{two}) can be written as
a combination of two one-soliton waves (\ref{onesoliton1}). The asymptotic
form of two-soliton solution in limits $t\rightarrow -\infty $ and $%
t\rightarrow \infty $ is similar to that of the one-soliton solution (\ref%
{onesoliton1}).

(i) Before collision (limit $t\rightarrow -\infty $).

(a) Soliton 1 ($\xi _{1}\approx 0$, $\xi _{2}\rightarrow -\infty $). 
\begin{equation}
\psi _{two}\rightarrow \tau _{1}\exp \left( i\Phi _{1}\right) \left[ \left(
1+Z_{1}\right) -\left( 1-Z_{1}\right) \tanh \frac{\xi _{1}}{2}\right] ,
\label{asym1a}
\end{equation}

(b) Soliton 2 ( $\xi _{2}\approx 0$, $\xi _{1}\rightarrow \infty $).%
\begin{equation}
\psi _{two}\rightarrow \tau _{1}Z_{1}\exp \left( i\Phi _{1}\right) \left[
\left( 1+Z_{2}\right) -\left( 1-Z_{2}\right) \tanh \frac{\xi _{2}+\delta _{0}%
}{2}\right] ,  \label{asym2a}
\end{equation}%
where the initial coordinate of the center of the solitary wave is removed
by $\delta _{0}=\ln A_{12}$.

(ii) After collision (limit $t\rightarrow \infty $).

(a) Soliton 1 ($\xi _{1}\approx 0$, $\xi _{2}\rightarrow \infty $).%
\begin{equation}
\psi _{two}\rightarrow Z_{2}\tau _{1}\exp \left( i\Phi _{1}\right) \left[
\left( 1+Z_{1}\right) -\left( 1-Z_{1}\right) \tanh \frac{\xi _{1}+\delta _{0}%
}{2}\right] ,  \label{asym1b}
\end{equation}

(b) Soliton 2 ( $\xi _{2}\approx 0$, $\xi _{1}\rightarrow -\infty $).%
\begin{equation}
\psi _{two}\rightarrow \tau _{1}\exp \left( i\Phi _{1}\right) \left[ \left(
1+Z_{2}\right) -\left( 1-Z_{2}\right) \tanh \frac{\xi _{1}}{2}\right] ,
\label{asym2b}
\end{equation}%
Analysis reveals that there is no amplitude exchange for soliton 1 and
soliton 2 during collision because of $\left\vert Z_{j}\right\vert =1$, $%
j=1,2$. However, from Eqs. (\ref{asym1a}) to (\ref{asym2b}) one can see that
there is a phase exchange $\delta _{0}/2$ for soliton 1 and soliton 2 during
collision. It shows that the information held in each soliton will almost
not be disturbed by each other in soliton propagation. These properties may
have potential application in future quantum communication.

\section{Conclusion}

Dark soliton dynamics of spinor BECs in an optical lattice is studied in
terms of a nonlinear schr\"{o}dinger equation by means of Hirota method.
Exact soliton solutions are obtained analytically and the elastic collision
of two solitons is demonstrated. It should be interesting to discuss how to
create and to detect such magnetic soliton in experiment. Using Landau-Zener
rf-sweeps at high fields (30 G) \cite{Miesner} a condensate was prepared in
the hyperfine state $\left\vert f=1,m_{f}=0\right\rangle $ of sodium, i.e.
the ground state of the spinor BECs. Then the atoms of the ground state can
be excited to the hyperfine state $\left\vert f=1,m_{f}=\pm 1\right\rangle $
by laser light experimentally. Therefore the excited state of the spinor
BECs, i.e. the magnetic soliton can be created. As the same discussion
before \cite{Lizd} the spatial-temporal spin variations in the soliton state
are significant. This makes it possible to make a direct detection of the
magnetic soliton of spinor BECs. By counting the difference numbers of the
population between the spin $+1$\ and $-1$\ Zeeman sublevel, the average of
spin component $<S^{z}>$\ is measured directly. Transverse components can be
measured by use of a short magnetic pulse to rotate the transverse spin
component to the longitudinal direction. Any optical or magnetic method
which can excite the internal transitions between the atomic Zeeman
sublevels can be used for this purpose. In current experiments in optical
lattices, the lattice number is in the range of $10$-$100$, and each lattice
site can accommodate a few thousand atoms. This leads to a requirement for
the frequency measurement precision of about 10-100 kHz. This is achievable
with current techniques.

The magnetic soliton of spinor BECs in an optical lattice is mainly caused
by the magnetic and the light-induced dipole-dipole interactions between
different lattice sites. Since these long-range interactions are highly
controllable the spinor BECs in optical lattice which is an exceedingly
clean system can serve as a test ground to study the static and dynamic
aspects of soliton excitations.

\section{Acknowledgement}

This work is supported by the Nature Science Foundation of China No.
10647122, the Doctoral Foundation of Education Bureau of Hebei Province of
China No. 2006110 and the key subject construction project of Heibei
Provincial University of China.

\end{document}